\documentclass[american,aps,pra,amsmath,amssymb,showpacs, twocolumn]{revtex4-1}
\usepackage[utf8]{inputenc}
\usepackage{amsmath}
\usepackage{amssymb}
\usepackage{graphicx}
\usepackage[percent]{overpic}
\usepackage{enumerate}
\usepackage{tikz,pgfplots}
\usepackage{mathbbol}

\newcommand{\down}{\ket{\hspace{-2.5pt}\downarrow}}
\newcommand{\up}{\ket{\hspace{-2.5pt}\uparrow}}

\global\long\def\ket#1{|#1\rangle}
\global\long\def\spst{|\psi_j\rangle}
\global\long\def\bra#1{\langle#1|}

\global\long\def\tr{\operatorname{tr}}
\global\long\def\dg{g}

\begin{document}

\title{Anticoherence measures for pure spin states}

\author{D.~Baguette and J.~Martin}
\affiliation{Institut de Physique Nucl\'eaire, Atomique et de
Spectroscopie, CESAM, Universit\'e de Li\`ege, B\^at.\ B15, B - 4000
Li\`ege, Belgium}

\begin{abstract}
The set of pure spin states with vanishing spin expectation value can be regarded as the set of the less coherent pure spin states. This set can be divided into a finite number of nested subsets on the basis of higher order moments of the spin operators. This subdivision relies on the notion of anticoherent spin state to order $t$: A spin state is said to be anticoherent to order $t$ if the moment of order $k$ of the spin components along any directions are equal for $k= 1, 2,\ldots, t$. Most spin states are neither coherent nor anticoherent, but can be arbitrary close to one or the other. In order to quantify the degree of anticoherence of pure spin states, we introduce the notion of anticoherence measures. By relying on the mapping between spin-$j$ states and symmetric states of $2j$ spin-$1/2$ (Majorana representation), we present a systematic way of constructing anticoherence measures to any order. We briefly discuss their connection with measures of quantum coherence. Finally, we illustrate our measures on various spin states and use them to investigate the problem of the existence of anticoherent spin states with degenerated Majorana points.
\end{abstract}


\date{July 12, 2017}

\maketitle

\section{Introduction} \label{sec:intro}

Quasi-classical or coherent states were first introduced for the quantum harmonic oscillator~\cite{Sch26}. They are the only states that both minimize the Heisenberg inequality for position and momentum and have equal dispersions on kinetic and potential energy. Along with these properties, coherent states lead to position and momentum expectation values with a time-dependence that has exactly the same form as their classical counterpart, which makes them the most classical states of a quantum harmonic oscillator~\cite{Gazeau,Gil90}. Their importance was widely recognized during the 1960’s, e.g.\ due to the work of Sudarshan~\cite{Sud63} and Glauber~\cite{Gla63} on the diagonal coherent states representation of the quantized electromagnetic field. Coherent states are by far not restricted to the harmonic oscillator and can be defined for a large variety of quantum systems~\cite{Gazeau}. In this paper, we are interested in quantum systems with arbitrary spin $j$, and, more generally, in any quantum system with three observables $J_i$ ($i=x,y,z$) satisfying the angular momentum commutation relations $[J_j,J_k]= i\epsilon_{jk\ell} J_\ell$ (with $\epsilon_{jk\ell}$ being the completely antisymmetric tensor and where we set $\hbar = 1$). Examples are multiphoton systems equipped with Stokes operators or atomic ensembles equipped with collective spin operators.
For systems of arbitrary spin $j$ (integer or half-integer), spin-$j$ coherent states are defined as the pure states for which the norm of the expectation value of the spin operator is \emph{maximal} and equal to $j$. More precisely, if $\mathbf{J}=(J_x,J_y,J_z)$ are the irreducible representations of dimension $2j+1$ of the spin operators, then for any spin-$j$ coherent state, we have that $\langle\mathbf{J}\rangle = j\mathbf{n}$ with $\mathbf{n}$ a real unit vector. Just as their classical counterparts, they are entirely characterized by a direction $\mathbf{n}$. Therefore, all spin-coherent states are connected to each other via a spin rotation. Apart from these quasi-classical spin states, there is a wealth of other spin states whose closeness to spin-coherent states can be quantified from the norm of their spin expectation value. In opposition to spin-coherent states, the less coherent spin states should be characterized by a zero spin expectation value. Such states have been studied in the literature, e.g.\ in the contexts of anticoherent spin states~\cite{Zimba_EJTP_3,Cra10,Bag15,Gir15,Per17}, completely entangled spin states~\cite{Kly02,Bag14} or multiphoton polarization states~\cite{Lui02,Kor06,Mar10,Lui13}. They also appear as some of the most non-classical spin states, where classicality of a spin state refers to the possibility of expressing it as a statistical mixture of spin-coherent states with positive weights~\cite{Gir08,Gir10}. Following Zimba~\cite{Zimba_EJTP_3}, we shall refer to states with vanishing spin expectation value as anticoherent states to order $1$. The general definition of anticoherence goes as follows: A pure spin-$j$ state $\ket{\psi_j}$ is said to be anticoherent to order $t$, or $t$-anticoherent, if $\langle({\mathbf{J}}\boldsymbol{\cdot}\mathbf{n})^k\rangle$ is independent of the unit vector $\mathbf{n}$ for $k = 1, \ldots, t$, where $\langle\cdot\rangle\equiv \bra{\psi_j}\cdot\ket{\psi_j}$. It readily follows from the definition that spin rotations preserve the order of anticoherence of a spin state and that any $t$-anticoherent state is necessarily $t'$-anticoherent for $t'< t$. In particular, all anticoherent states are $1$-anticoherent and thus are among the less coherent states. 

Most spin states are neither coherent nor anticoherent, but can be arbitrary close to one or the other. In particular, a slight modification of a state can be sufficient to make it loose its coherent or anticoherent character. Still, most of the state's physical properties would be slightly perturbed and the state could be used for the same practical purposes as the original one. The main goal of this work is therefore to introduce measures of anticoherence to position any state between the two extreme sets of coherent and anticoherent spin states. Our approach bears some analogy with the design in~\cite{Gir10} of the measure of quantumness for spin states, or the proposals in~\cite{Mar10} for measures of quantum degrees of polarization for multi-photon states. The paper is organized as follows. In Sec.~\ref{sec:acs_properties}, we give some examples of anticoherent spin states, and review some of their properties and their characterization in the Majorana representation. In Sec.~\ref{sec:ac_measures}, we propose an axiomatic approach to the definition of measures of anticoherence to any order. We then elaborate several measures of anticoherence and introduce a systematic way to construct such measures based on operator distances. In Sec.~\ref{sec:applications}, we use our measures of anticoherence to study the existence of anticoherent states for spin quantum numbers up to $j=10$.

\section{Anticoherent states: examples, characterization and properties}
\label{sec:acs_properties}

\subsection{Examples and properties}
A paradigmatic example of $1$-anticoherent spin-$j$ state is Schrödinger's cat state
\begin{equation}\label{Scatstates}
\ket{\psi_j^{\mathrm{cat}}}=\frac{\ket{j,-j}+\ket{j,j}}{\sqrt{2}},
\end{equation}
written here in the standard basis $\{\ket{j,m}:m=-j,-j+1,\ldots,j\}$ formed by the common eigenstates of $\mathbf{J}^2$ and $J_z$ of eigenvalues $j(j+1)$ and $m$ respectively. For any integer or half-integer $j>1/2$, the states (\ref{Scatstates}) are characterized by $\langle {\mathbf{J}}\boldsymbol{\cdot}\mathbf{n}\rangle = 0$ for any $\mathbf{n}$, $\langle J_x^2\rangle=j/2$, $\langle J_y^2\rangle=j/2(1-\delta_{j,1})$ and $\langle J_z^2\rangle=j^2$, from which it follows that they are $1$-anticoherent but never $2$-anticoherent because all $\langle J_i^2\rangle$ for $i=x,y,z$ cannot be equal for $j>1/2$. Another example of $1$-anticoherent (but not $2$-anticoherent) state for any integer $j$ is the Dicke state $\ket{j,0}$. 

As for $2$-anticoherent states, a first example is given by the spin-2 state \cite{footnote1}
\begin{equation}\label{tetrahedron}
\ket{\psi_2^{\mathrm{tet}}}=\frac{1}{2}(\ket{2,-2}+ i \sqrt{2}\, \ket{2,0} + \ket{2,2}),
\end{equation}
for which a direct calculation yields
\begin{equation}\label{expvaltetra}
\begin{aligned}
& \langle {\mathbf{J}}\boldsymbol{\cdot}\mathbf{n}_1 \rangle = 0,\\[4pt]
& \langle({\mathbf{J}}\boldsymbol{\cdot}\mathbf{n}_1)({\mathbf{J}}\boldsymbol{\cdot}\mathbf{n}_2)\rangle = 2\,\mathbf{n}_1\boldsymbol{\cdot}\mathbf{n}_2,
\end{aligned}
\end{equation}
for any orientations $\mathbf{n}_1,\mathbf{n}_2$. This shows that the state (\ref{tetrahedron}) is indeed $2$-anticoherent as (\ref{expvaltetra}) implies that $\langle {\mathbf{J}}\boldsymbol{\cdot}\mathbf{n}_1 \rangle$ and $\langle({\mathbf{J}}\boldsymbol{\cdot}\mathbf{n}_1)^2\rangle$ do not depend on $\mathbf{n}_1$. As the expectation values (\ref{expvaltetra}) are both invariant under rotations, the measurement results of the product of at most two spin operators will not depend on the orientation  of the spin system. In other words, no experiments relying on the measurement of the product of at most two spin operators will allow us to determine whether the system has been rotated or not. Surprisingly, the transition probability between a spin-$2$ state and the state obtained from it by a rotation has been shown to be minimized by (\ref{tetrahedron}) for a large range of angles, making it an optimal state in detecting rotations~\cite{Chr17}. The state (\ref{tetrahedron}) was also shown to be optimal for reference frame alignment~\cite{Kol08}.

More generally, for any $t$-anticoherent spin state, the expectation value of the product of $t'\leqslant t$ spin operators is invariant under rotation and given by the value~\cite{BagMar}
\begin{equation}\label{eq:meanval}
\langle({\mathbf{J}}\boldsymbol{\cdot}\mathbf{n}_1)\ldots ({\mathbf{J}}\boldsymbol{\cdot}\mathbf{n}_{t'})\rangle = \frac{\tr \left[ ({\mathbf{J}}\boldsymbol{\cdot}\mathbf{n}_1)\ldots ({\mathbf{J}}\boldsymbol{\cdot}\mathbf{n}_{t'}) \right]}{2j+1}
\end{equation}
that only depends on $\mathbf{n}_1,\ldots,\mathbf{n}_{t'}$ and $j$. In particular, for $t'=1$ and $t'=2$, Eq.~(\ref{eq:meanval}) reduces to Eq.~(\ref{expvaltetra}). This time, Eq.~(\ref{eq:meanval}) implies that no experiments relying on the measurement of homogeneous functions of the spin operators up to degree $t$ will allow to determine whether a spin system in a $t$-anticoherent state has been rotated or not. 

However, anticoherent states to arbitrary order do not necessarily exist in a spin system with given spin quantum number $j$. For instance, no pure spin-$j$ anticoherent states of order $t>j$ do exist~\cite{Bag14}. Nevertheless, it has been shown that pure anticoherent states to any order $t$ exist provided that $j$ is sufficiently large, typically $j\sim t^2$~\cite{Bag15}. While spin rotations preserve the order of anticoherence of a spin state, all anticoherent states to a given order are not \emph{necessarily} connected by rotations, as is the case for coherent states. For instance, for $j=2$, it has been shown in~\cite{Bag14} that there is an infinite number of $1$-anticoherent states of the form
\begin{equation}\label{eq:psimu}
\ket{\psi_2^\mu}=\frac{1}{\sqrt{2+|\mu|^2}} ( \ket{2,-2} + \mu \ket{2,0} + \ket{2,2}),
\end{equation}
with $\mu\in\mathbb{C}$, that are not connected by rotations, whereas for $j=1$ and $j=3/2$, all $1$-anticoherent states are connected by rotations. The states (\ref{eq:psimu}) form a linear subspace spanned by the two $1$-anticoherent states $\ket{\psi_2^{\mathrm{cat}}}$ and $\ket{2,0}$. The concept of anticoherent subspaces has been developed and studied in~\cite{Per17}.

A characterization of anticoherent spin states can be given in terms of total variance.  The total variance of a pure spin-$j$ state $|\psi_j\rangle$ is defined as~\cite{Kly03,Kly07,Kly07b,Saw12}
\begin{equation}\label{eq:TVspin}
\begin{aligned}
\mathbb{V}(|\psi_j\rangle) &= \sum_{i=x,y,z} (\langle J_i^2 \rangle-\langle J_i \rangle^2)= j(j+1)-|\langle \mathbf{J}\rangle|^2
\end{aligned}
\end{equation}
and is a measure of the overall level of quantum fluctuations of the spin in state $|\psi_j\rangle$. It is invariant under rotations, minimal for spin-coherent states ($\mathbb{V}=j$) and maximal whenever the spin expectation vanishes (in which case $\mathbb{V}=j(j+1)$)~\cite{Kly03}; hence, it is maximal for anticoherent states. The total variance has proved a useful tool in different contexts such as entanglement quantification~\cite{Ill10,Leu14}, entanglement classification under stochastic local operations and classical communication (SLOCC)~\cite{Saw12}, and control of coherence~\cite{Oli13}.

\subsection{Majorana representation of spin-$j$ states}
\label{subsec:Majo}

In this subsection, we introduce the Majorana representation for spin systems that maps spin-$j$ states to $2j$ spin-$1/2$ symmetric states. This representation has been widely used to deal with various problems about spin states~\cite{Maj,Gir10,Cra10,Mar12,Bag15,Chr17}. We then review the conditions for $t$-anticoherence in the light of this mapping in terms of reduced density matrices.

\subsubsection{Spin-$j$ states and rotations in the Majorana representation}

Any spin-$j$ state can be expanded in the standard angular momentum basis as
\begin{equation}\label{eq:jm_decomp}
\spst = \sum_{m=-j}^j c_{m}\, \ket{j,m}
\end{equation}
with $c_m\in\mathbb{C}$ and $\sum_{m=-j}^j |c_{m}|^2=1$. In his seminal paper on the variation of orientation of atoms propagating in a variable magnetic field~\cite{Majorana}, Ettore~Majorana introduced another representation of spin-$j$ states based on a one-to-one correspondence ($\leftrightarrow $) with symmetric states of $2j$ spin-$1/2$,
\begin{align}
&\ket{j,m}\leftrightarrow \ket{D_{2j}^{(j-m)}}\\
&\spst \leftrightarrow \ket{\psi_S} = \sum_{k=0}^{2j} c_{j-k}\,\ket{D_{2j}^{(j-m)}}\label{symMaj}
\end{align}
with $\ket{D_{2j}^{(j-m)}}$ being the symmetric Dicke states defined as 
\begin{equation}\label{eq:mapping}
\ket{D_{2j}^{(j-m)}}=\mathcal{N}\sum_{\pi}\underbrace{\ket{\hspace{-2pt}\downarrow}\otimes\ldots \otimes\ket{\hspace{-2pt}\downarrow}}_{j-m}\otimes\underbrace{\ket{\hspace{-2pt}\uparrow}\otimes\ldots \otimes\ket{\hspace{-2pt}\uparrow}}_{j+m}
\end{equation}
where $\mathcal{N}$ is a normalization constant, and the sum runs over the $(2j)!$ permutations $\pi$ of the $j-m$ spin $\ket{\hspace{-2pt}\downarrow}\equiv \ket{\tfrac{1}{2}-\tfrac{1}{2}}$ and the $j+m$ spin $\ket{\hspace{-2pt}\uparrow}\equiv \ket{\tfrac{1}{2},\tfrac{1}{2}}$. 
The symmetric state $\ket{\psi_S}$ in Eq.~(\ref{symMaj}) can also be written in the form (\ref{eq:mapping}) as
\begin{equation}\label{spinjMaj}
\ket{\psi_S} =\mathcal{N}\sum_{\pi}\ket{\phi_{\pi(1)}}\otimes \ldots \otimes \ket{\phi_{\pi(N)}}
\end{equation}
where, for $k=0,\ldots,2j$,
\begin{equation}\label{coh}
\ket{\phi_{k}}=\cos(\tfrac{\theta_k}{2}) \ket{\tfrac12,\tfrac12}+\sin(\tfrac{\theta_k}{2})e^{i\varphi_k} \ket{\tfrac12,-\tfrac12}
\end{equation}
is a spin-$1/2$ state parametrized by the angles $(\theta_k,\varphi_k)$ specifying a point on the Bloch sphere. Thus, in the Majorana representation, a spin-$j$ state is fully specified by $2j$ points (Majorana points) on the Bloch sphere. For spin-$j$ coherent states, all Majorana points are located at the same position on the Bloch sphere ($2j$-fold degenerated Majorana point) and the corresponding symmetric state is \emph{separable}. The standard basis states $\ket{j,m}$ correspond to $(j-m)$ Majorana points at the south pole and $(j+m)$ Majorana points at the north pole of the Bloch sphere. 

\paragraph{Rotations} The rotation of a spin-$j$ state of an angle $\theta$ around the axis $\mathbf{n}$ is represented by the unitary operator $R_{\mathbf{n}}(\theta)=\exp({-i\theta\mathbf{J}\boldsymbol{\cdot}\mathbf{n}})$. In the Majorana representation, it is equivalent to the individual rotations of all spin-$1/2$ of the same angle $\theta$ around the same axis $\mathbf{n}$, represented by the symmetric local unitary (LU) operator $r_\mathbf{n} \otimes \ldots \otimes r_\mathbf{n}$, where $r_\mathbf{n}=\exp(-i \theta\, \boldsymbol{\sigma}\boldsymbol{\cdot}\mathbf{n}/2)$ acts on a single spin-$1/2$, with $\boldsymbol{\sigma}=(\sigma_x,\sigma_y,\sigma_z)$ being the vector of Pauli matrices. This, in turn, corresponds to a rigid rotation of all Majorana points on the Bloch sphere.

\subsubsection{Anticoherence in the Majorana representation}

An elegant necessary and sufficient condition for $t$-anticoherence has been derived using the Majorana representation~\cite{Gir15}. This condition is expressed in terms of the $t$ spin-$1/2$ reduced density matrices of the $2j$ spin-$1/2$ state (\ref{symMaj}). As the state (\ref{symMaj}) is invariant under permutation of the spins, all its $t$ spin-$1/2$ reduced density operators are equal and also invariant under permutation of the spins. Denoting by $\rho_t$ these density operators or any of their matrix representation in a basis spanning the symmetric subspace of $t$ spin-$1/2$, the following equivalence holds~\cite{Gir15}
\begin{equation}\label{eq:defacrhot}
\begin{gathered}
\spst \text{ is $t$-anticoherent}\\
\Updownarrow\\
\rho_t = \tr_{1\ldots N-t}(\ket{\psi_S}\bra{\psi_S})= \frac{\mathbb{1}_{t+1}}{t+1}
\end{gathered}
\end{equation}
where $\mathbb{1}_{t+1}=\sum_{k=0}^{t}\ket{D_t^{(k)}}\bra{D_t^{(k)}}$ is the identity operator in the symmetric subspace of dimension $t+1$. In contrast, spin coherent states, which are in one-to-one correspondence with \emph{pure separable} symmetric states, are characterized by \emph{pure} (i.e.\ rank $1$) reduced density operators. In all generality, the reduced density operator $\rho_t$ of the state (\ref{symMaj}) has the compact expression (see Appendix A of \cite{Bag14})
\begin{equation}\label{dop}
\rho_t=\sum_{k_1=0}^{t}\sum_{k_2=0}^{t}(\rho_t)_{k_1 k_2} \ket{D_t^{(k_1)}}\bra{D_t^{(k_2)}},
\end{equation}
with the matrix elements in the Dicke basis $\{\ket{D_t^{(k)}}:0\leqslant k \leqslant t\}$
\begin{equation}\label{eq:vtqvtl}
(\rho_t)_{k_1 k_2}=\sum_{k=0}^{N-t}c_{j-k-k_1}\, c_{j-k-k_2}^{*}\,\Gamma_k^{k_1k_2},
\end{equation}
where
\begin{equation}
\label{Gamma}
\Gamma_k^{k_1k_2}=\frac{1}{C_{2j}^{t}}\sqrt{C_{k+k_1}^{k}C^{t-k_1}_{2j-k-k_1}C_{k+k_2}^{k}C^{t-k_2}_{2j-k-k_2}}
\end{equation}
and ${C_{q}^{\ell}} = \binom{q}{\ell}$ if $0\leqslant \ell \leqslant q$ and 0 otherwise.

\section{Measures of anticoherence}\label{sec:ac_measures}

In this section, we present our abstract definition of measures of $t$-anticoherence which consists of a list of conditions that every measures must satisfy. Using the tools presented in Sec.~\ref{sec:acs_properties}, we then explicitly construct several measures of anticoherence, and discuss their relation to measures of quantum coherence.

\subsection{Axiomatic definition of measures of anticoherence for pure spin states}

Let $\spst$ be a pure spin-$j$ state and $t$ be a positive integer such that $t< 2j$. We define a measure of anticoherence to order $t$ (or $t$-anticoherence measure) for pure spin-$j$ states as a positive function $\mathcal{A}_t(\spst)$ satisfying the \emph{minimal} set of conditions: 
\begin{enumerate}[i.]
\item $\mathcal{A}_t(\spst)=0$ $\;\Leftrightarrow\;$ $\spst$ is coherent. \label{ax:coherent}\label{ax:first}
\item $\mathcal{A}_t(\spst)=1$ $\;\Leftrightarrow\;$ $\spst$ is $t$-anticoherent. \label{ax:anticoherent}
\item $\mathcal{A}_t(\spst) \in [0,1]$ for any $\spst$.\label{ax:interval}\label{ax:range}
\item $\mathcal{A}_t(\spst)$ is invariant under global phase changes and arbitrary spin rotations. \label{ax:rotation}\label{ax:last}
\end{enumerate}

The first three conditions ensure that coherent states, respectively anticoherent states to order $t$, are the only states minimizing, respectively maximizing, any measures of anticoherence to order $t$.
The last condition ensures that the value taken by measures of anticoherence does not depend on a particular coordinate system. It is equivalent to the equality
\begin{equation}
\mathcal{A}_t\big(\spst\big) = \mathcal{A}_t\big(e^{i\alpha} R_{\mathbf{n}}(\theta)\spst\big)\quad \forall\,\spst
\end{equation}
for any $\theta,\alpha\in\mathbb{R}$ and $\mathbf{n}\in\mathbb{R}^3$. 

Let us mention some of the direct implications of the conditions \ref{ax:first}-\ref{ax:last} and known properties of anticoherent states reviewed in Secs.~\ref{sec:intro} and \ref{sec:acs_properties}. First, as any $t$-anticoherent state is also $t'$-anticoherent for $t'\leqslant t$, we have the following relationship between measures of anticoherence to different orders
\begin{equation}
\mathcal{A}_{t}(\spst)=1 \quad\Rightarrow\quad \mathcal{A}_{t'}(\spst)=1\qquad\forall \, t'\leqslant t.
\end{equation}
Second, as any measure of anticoherence vanishes only for coherent states, we have that
\begin{equation}
\begin{aligned}
\mathcal{A}_{t}(\spst)=0 \quad\Leftrightarrow\quad \mathcal{A}_{t'}(\spst)=0\qquad\forall \, t'<2j,\\[3pt]
\mathcal{A}_{t}(\spst)>0 \quad\Leftrightarrow\quad \mathcal{A}_{t'}(\spst)>0\qquad\forall \, t'<2j.
\end{aligned}
\end{equation}
Third, as no pure spin-$j$ anticoherent states of order $t>j$ exist, we have that
\begin{equation}
\mathcal{A}_{t}(\spst)<1 \qquad\forall \, t>j.
\end{equation}

Because the equivalence (\ref{eq:defacrhot}) for $t$-anticoherence is expressed in terms of $t$ spin-$1/2$ reduced density matrices, we consider $t<2j$ for practical purposes. In the next Subsections, we explicitly construct several measures of $t$-anticoherence for any order $t$.

\subsection{Measure of $1$-anticoherence based on total variance}

The total variance (\ref{eq:TVspin}) can be used to construct a measure of $1$-anticoherence, that we define by
\begin{equation}\label{acm_tv}
\mathcal{A}_1^{\mathbb{V}}(|\psi_j\rangle)=\frac{\mathbb{V}(|\psi_j\rangle)-j}{j^2}= \frac{j^2-|\langle \mathbf{J}\rangle|^2}{j^2}.
\end{equation}
Equation (\ref{acm_tv}) is indeed a measure of $1$-anticoherence as it satisfies all conditions \ref{ax:first}--\ref{ax:last}. It depends linearly on $\mathbb{V}$ but other real functions of $\mathbb{V}$ could also be used to define other equally valid measures of $1$-anticoherence. As an illustration, the measure of $1$-anticoherence (\ref{acm_tv}) for the Schrödinger cat state (\ref{Scatstates}) and the Dicke state $\ket{j,-j+1}$ is given, for all $j>1/2$, by
\begin{equation}
\begin{aligned}
&\mathcal{A}_1^{\mathbb{V}}(\ket{\psi_j^{\mathrm{cat}}}) = 1,  \\
&\mathcal{A}_1^{\mathbb{V}}(\ket{j,-j+1}) = \frac{2j-1}{j^2}.
\end{aligned}
\end{equation}
This shows that the Schrödinger cat state is $1$-anticoherent for all $j>1/2$, whereas the Dicke state $\ket{j,-j+1}$ is never $1$-anticoherent.

In the next section, we show how to generalize to any $t>1$ the measure (\ref{acm_tv}) based on the total variance. While the total variance involves second order moments of the spin operators, its generalizations are based on higher moments of the spin operators. They will enable us to characterize further $1$-anticoherent states that have the same total variance but are not necessarily connected by a rotation.

\subsection{Measures of anticoherence based on purity}

Let us denote by $\lambda_1,\ldots,\lambda_{t+1}$ the eigenvalues of the reduced density operator $\rho_t=\tr_{1\ldots N-t}(\ket{\psi_S}\bra{\psi_S})$ where $\ket{\psi_S}$ is in one-to-one correspondence with $\spst$ [see Eq.~(\ref{symMaj})]. The purity of $\rho_t$, for any $t$,
\begin{equation}\label{purity}
R_t(\spst)\equiv\tr(\rho_t^2)=\sum_{i=1}^{t+1}\lambda_i^2,
\end{equation}
can be used to form a simple measure of $t$-anticoherence, that we define as
\begin{equation}\label{eq_acm_pure}
\mathcal{A}_t^{R}(\spst)= \frac{t+1}{t}\left[1-R_t(\spst)\right].
\end{equation}
This measure is the rescaled linear entropy $S_L=1-R_t$ of the reduced state $\rho_t$ so that $\mathcal{A}_t^{R}\in[0,1]$. The von Neumann entropy $S=-\mathrm{tr}(\rho_t \ln\rho_t)$ could also be used to form a similar measure of anticoherence based on the bipartite entanglement between $t$ and $2j-t$ spin-$1/2$. The linear entropy is invariant under (symmetric) LU, maximal only for maximally mixed states and vanishes only for pure states. But anticoherent spin states are precisely in one-to-one correspondence with symmetric states having maximally mixed reduced states $\rho_t$, while coherent states are  in one-to-one correspondence with symmetric states having pure reduced states $\rho_t$. Hence, Eq.~(\ref{eq_acm_pure}) satisfies all conditions i-iv for a proper measure of $t$-anticoherence according to our definition. Inserting Eq.~(\ref{dop}) with (\ref{eq:vtqvtl}) into (\ref{purity}), the expression for the purity of $\rho_t$ in terms of expansion coefficients of $|\psi_j\rangle$ in the standard basis (see Eq.~(\ref{eq:jm_decomp})) follows,
\begin{equation}\label{Rtcm}
\begin{aligned}
R_t(\spst) = \sum_{k_1=0}^t\sum_{k_2=0}^t &\Bigg| \sum_{k=-j}^{j-t} c^*_{k+k_1} c_{k+k_2} \Gamma_{j+k}^{k_1k_2} \Bigg|^2,
\end{aligned}
\end{equation}
where $\Gamma_{k}^{k_1k_2}$ is given by Eq.~(\ref{Gamma}). Therefore, the purity and the purity-based measure of anticoherence~(\ref{eq_acm_pure}) are straightforward to compute once the expansion of the state $\spst$ in the standard basis is known. 

The representation (\ref{symMaj}) of spin-$j$ states used to arrive at this result is a convenient theoretical tool but the $2j$ spin-$1/2$ making up a total spin $j$ might be purely fictitious and, if they exist, might be individually inaccessible in an experiment. Therefore, it is very relevant to express the measure (\ref{eq_acm_pure}) solely in terms of spin-$j$ expectation values. This can be done by using the tensor representation of spin states introduced in~\cite{Gir15}. This representation relies on so-called Weinberg matrices, $S_{\mu_1\mu_2\ldots\mu_{2j}}$ where $\mu_i \in \{0,x,y,z\}$, that form an overcomplete basis for density matrices of general spin-$j$ states. They can be defined from the operators~\cite{Gir15}
\begin{equation}\label{Sop}
S_{\mu_1 \mu_2 \ldots \mu_{2j}}=\mathcal{P}_S\,(\sigma_{\mu_1} \otimes \sigma_{\mu_2}\otimes \ldots \otimes \sigma_{\mu_{2j}})\,\mathcal{P}_S^\dagger
\end{equation}
where $\sigma_0$ is the identity operator and $\sigma_i$ for $i=x,y,z$ are the Pauli operators acting on a single spin-$1/2$, and $\mathcal{P}_S$ is the projector from the Hilbert space $\mathcal{H}$ of $2j$ spin-$1/2$ of dimension $2^{2j}$ onto its symmetric subspace $\mathcal{S}$ of dimension $2j+1$. The operators (\ref{Sop}) can be represented, in the symmetric Dicke basis, by square matrices of dimension $2j+1$ (Weinberg matrices). The operators $\sigma_{\mu_1} \otimes \sigma_{\mu_2}\otimes \ldots \otimes \sigma_{\mu_{2j}}$ can be represented, in the computational basis of $2j$ spin-$1/2$, by square matrices of dimension $2^{2j}$. Finally, the projector $\mathcal{P}_S$ can be represented by a rectangular matrix of dimension $(2j+1)\times (2^{2j})$, mapping $\mathcal{H}$ onto $\mathcal{S}$. Note that because of the symmetrization, the order of the subscripts $\mu_1,\ldots,\mu_{2j}$ of the Weinberg matrices is irrelevant. Any spin-$j$ density matrix $\rho_j$ can be expressed as
\begin{equation}\label{Weindecomp}
\rho_j = \frac{1}{4^j}\sum_{\mu_1, \ldots,\mu_{2j}}\langle S_{\mu_1 \mu_2 \ldots \mu_{2j}}\rangle S_{\mu_1 \mu_2 \ldots \mu_{2j}}
\end{equation}
where the sum is over $0,x,y,z$ for each of the $\mu_i$ and with $\langle S_{\mu_1 \mu_2 \ldots \mu_{2j}}\rangle=\mathrm{Tr}(\rho_j S_{\mu_1 \mu_2 \ldots \mu_{2j}})$. One of the advantages in using this representation is the simple expression of the $t$ spin-$1/2$ reduced density matrices~\cite{Gir15}, 
\begin{equation}
\rho_t = \frac{1}{2^{t}} \sum_{\mu_1, \ldots, \mu_t}\langle S_{\mu_1 \ldots \mu_t 0 \ldots 0}\rangle S_{\mu_1 \ldots \mu_t}.
\end{equation}
In particular, the purity of $\rho_t$ [Eq.~(\ref{purity})] is given by~\cite{Gir15}
\begin{equation}\label{eq:trrho2}
R_t(\spst)=\tr(\rho_t^2) = \frac{1}{2^{t}} \sum_{\mu_1, \ldots, \mu_{t}}  \langle S_{\mu_1 \ldots \mu_t 0 \ldots 0}\rangle^2
\end{equation}
with $\langle S_{\mu_1 \ldots \mu_t 0 \ldots 0}\rangle=\bra{\psi_j} S_{\mu_1 \ldots \mu_t 0 \ldots 0}\ket{\psi_j}$. Now, to have the measure (\ref{eq_acm_pure}) solely in terms of spin-$j$ expectation values, it remains to express the Weinberg matrices $S_{\mu_1 \ldots \mu_t 0 \ldots 0}$ in terms of the spin operators $J_x, J_y, J_z$ and the identity operator $J_0$. A general procedure is presented in~\cite{Gir15}. Let us illustrate the method for the measures of $1$- and $2$-anticoherence. For any spin quantum number $j>1$, we have
\begin{equation} \label{eq:S0}
S_{0\ldots0}=J_0,\qquad S_{a0\ldots0}=\frac{J_a}{j}
\end{equation}
and
\begin{equation}\label{eq:Sab}
S_{ab0\ldots0}=\frac{1}{(2j-1)}\left(\frac{J_a J_b + J_b J_a}{j} - \delta_{ab}J_0\right)
\end{equation}
with $a,b=x,y,z$. By combining Eqs.~(\ref{eq:S0}) and (\ref{eq:Sab}) with (\ref{eq:trrho2}), we get
\begin{align}
R_1(\spst) = {}& \frac{1}{2}\left(1+\frac{|\langle \mathbf{J}\rangle|^2}{j^2}\right),\label{R1J}\\
R_2(\spst) ={}& \frac{1}{4}+\frac{|\langle \mathbf{J}\rangle|^2}{2j}+\sum_{a,b} \left( \frac{\langle J_a J_b + J_b J_a \rangle /j -\delta_{ab} }{2(2j-1)}\right)^2.\label{R2J}
\end{align}
Using Eq.~(\ref{R1J}), we find that the anticoherence measure (\ref{eq_acm_pure}) for $t=1$ coincide with the measure (\ref{acm_tv}) based on the total variance, 
\begin{equation}
\mathcal{A}_1^R(\spst) = \mathcal{A}_1^{\mathbb{V}}(\spst).
\end{equation}
Hence, the anticoherence measures (\ref{eq_acm_pure}) for $t>1$ provide a simple generalization of (\ref{acm_tv}). In particular, using Eq.~(\ref{R2J}), Eq.~(\ref{eq_acm_pure}) for $t=2$ can be written, using the angular momentum commutation relations and after some algebra, as
\begin{equation}
 \mathcal{A}_2^R(\spst) = \frac{\mathbb{W} +\alpha}{\beta},
\end{equation}
with
\begin{equation}\label{W}
\mathbb{W} = \mathbb{V} - \frac{1}{2j (j-1)} \sum_{a,b} \langle J_a J_b \rangle \langle J_b J_a \rangle,
\end{equation}
and
\begin{equation}
\alpha = \frac{j (j^2-2j+3) }{2 (j-1)},\quad \beta =\frac{(2j-1)^2 j}{3 (j -1)}.
\end{equation}
The quantity $\mathbb{W}$ defined in Eq.~(\ref{W}) for $j>1$, involving correlators of two spin operators, is minimal for coherent states ($\mathbb{W}=-\alpha<0$) and maximal for anticoherent states to order $2$ ($\mathbb{W}=\beta-\alpha>0$). Hence, $\mathbb{W}$ (or any linear function of it) can be viewed as a generalization to order 2 of the total variance.

\subsection{Measures of anticoherence based on operator distances}

Let $d(\rho,\sigma)$ be a distance between any two density operators $\rho$ and $\sigma$ with the property of invariance under unitary transformation, i.e.\ $d(\rho,\sigma)=d(U \rho U^\dagger, U \sigma U^\dagger)$ with $U$ any unitary transformation. We also assume that the maximal distance to the maximally mixed state is achieved only for pure states, i.e.\ rank-$1$ density operators. In particular, this holds for the most commonly used distances, such as all distances induced by Schatten-$p$ norms and the Bures distance~\cite{Zyczkowski_book}. For any such distance, we define a measure of $t$-anticoherence as
\begin{equation}
\begin{aligned}\label{eq:ac_meas_dist}
\mathcal{A}^d_t(\spst) ={}& 1 - K_t^{-1}\, d\left(\rho_t,\frac{\mathbb{1}_{t+1}}{t+1}\right)
\end{aligned}
\end{equation}
where $K_t$ is the distance between any pure state and the maximally mixed state.
Using the unitarily invariance of the distance and the fact that the maximally mixed state does not change under unitary transformations, Eq.~(\ref{eq:ac_meas_dist}) can be written as
\begin{equation}
\begin{aligned}\label{eq:ac_meas_dist2}
\mathcal{A}^d_t(\spst) ={}&1 - K_t^{-1}\, d\left(\mathrm{diag}(\lambda_1,\ldots,\lambda_{t+1}),\frac{\mathbb{1}_{t+1}}{t+1}\right)
\end{aligned}
\end{equation}
with
\begin{equation}\label{Kt}
K_t = d\left(\mathrm{diag}(1,0,\ldots,0),\frac{\mathbb{1}_{t+1}}{t+1}\right),
\end{equation}
where $\lambda_1,\ldots,\lambda_{t+1}$ are the eigenvalues of $\rho_t$.
The measure (\ref{eq:ac_meas_dist}) is invariant under (symmetric) LU because the distance is unitarily invariant. It is minimal if and only if (iff) the state $\spst$ is coherent. Indeed, in this case, the reduced density matrix $\rho_t$ is pure and thus unitarily equivalent to $\mathrm{diag}(1,0,\ldots,0)$, so that the distance in Eq.~(\ref{eq:ac_meas_dist2}) is maximal and equal to $K_t$, leading to $\mathcal{A}^d_t(\spst)=0$. It is maximal iff the state $\spst$ is $t$-anticoherent, because the distance in Eq.~(\ref{eq:ac_meas_dist2}) is minimal and equal to $0$ iff the reduced density matrix $\rho_t$ is maximally mixed, in which case $\mathcal{A}^d_t(\spst)=1$. Therefore, Eq.~(\ref{eq:ac_meas_dist}) satisfies all conditions \ref{ax:first}--\ref{ax:last} for a measure of $t$-anticoherence.

Let us now exemplify this construction on different operator distances.
For the Hilbert-Schmidt distance $d^\mathrm{HS}(\rho,\sigma)=\sqrt{\mathrm{tr}[(\rho-\sigma)^2]}$, the constant (\ref{Kt}) is equal to $K_t = \sqrt{t/(t+1)}$, and the Hilbert-Schmidt measure of $t$-anticoherence reads
\begin{equation}\label{actHS}
\mathcal{A}_t^{\mathrm{HS}}(\spst)= 1-\sqrt{\frac{t+1}{t} \sum_{i=1}^{t+1} \left( \lambda_i - \frac{1}{t+1} \right)^2}.
\end{equation}
This measure is related to the one based on the purity of $\rho_t$ given in Eq.~(\ref{eq_acm_pure}) as we have
\begin{equation}
\mathcal{A}_t^{\mathrm{HS}} = 1-\sqrt{1-\mathcal{A}_t^R}.
\end{equation} 
For the trace distance $d^\mathrm{tr}(\rho,\sigma)=\mathrm{tr}[\sqrt{(\rho-\sigma)^2}]/2$, the constant (\ref{Kt}) is equal to $K_t = t/(t+1)$, and the trace measure of $t$-anticoherence reads
\begin{equation}\label{acttr}
\mathcal{A}_t^{\mathrm{tr}}(\spst)=1-\frac{t+1}{2t}\sum_{i=1}^{t+1}\left| \lambda_i - \frac{1}{t+1} \right|.
\end{equation}
The Bures distance  $d^\mathrm{Bures}(\rho,\sigma)=\sqrt{2-2 F(\rho, \sigma)}$ given in terms of the fidelity between the states $\rho$ and $\sigma$, $F(\rho, \sigma)=\mathrm{tr}(\sqrt{\sqrt{\rho} \sigma \sqrt{\rho}})$, is an example of unitarily invariant distance which is not induced by a norm. In this case, the constant (\ref{Kt}) is equal to $K_t = \sqrt{2 (1-1/\sqrt{t+1})}$, and the Bures measure of $t$-anticoherence reads
\begin{equation}\label{actBures}
\mathcal{A}_t^{\mathrm{Bures}}(\spst)=1-\sqrt{\frac{\sqrt{t+1}-\sum_{i=1}^{t+1}\sqrt{\lambda_i}}{\sqrt{t+1}-1}}.
\end{equation}
Extensive numerical computations seem to indicate that the Bures measure of anticoherence  is monotonous in $t$, that is 
\begin{equation}\label{eq:decreasing}
\mathcal{A}_{t}^{\mathrm{Bures}}(\spst) \geqslant \mathcal{A}_{t+1}^{\mathrm{Bures}}(\spst)\quad \forall\,t,
\end{equation}
for any spin-$j$ state $\spst$. Note that this inequality does not follow only from the contractivity property of the Bures distance. However, violation of this inequality for $t=1$ has been observed for the Hilbert-Schmidt and trace distances, in particular for the spin-$5/2$ state
\begin{equation}\label{eq:psie}
\ket{\psi_{5/2}}=\frac{1}{2}\left(\ket{\tfrac{5}{2},-\tfrac{5}{2}}+\ket{\tfrac{5}{2},-\tfrac{3}{2}}+\ket{\tfrac{5}{2},\tfrac{3}{2}}+\ket{\tfrac{5}{2},\tfrac{5}{2}}\right).
\end{equation}

\subsection{Measures of anticoherence vs measures of coherence}

The measures of anticoherence introduced in this work are related to the notions of spin-coherent and spin-anticoherent states. These notions seem \emph{a priori} disconnected from those of measures of quantum coherence, aimed at quantifying the importance of a density matrix’ off-diagonal entries in a specified basis~\cite{Ple14}. Yet, an explicit connection can be made, as we now  explain. Let us remember that $\rho_t$ denotes the $t$ spin-$1/2$ reduced density operator $\tr_{1\ldots N-t}(\ket{\psi_S}\bra{\psi_S})$ expressed in the Dicke basis, spanning the symmetric subspace $\mathcal{S}_t$ of $t$ spin-$1/2$, as a $(t+1)\times (t+1)$ matrix. Then, Theorem 2 of Ref.~\cite{Bru16} implies that for any distance-based measure of $t$-anticoherence $\mathcal{A}_{t}^d$ with contractive distance $d$, $1-\mathcal{A}_{t}^d$ is directly proportional to the maximal coherence of $\rho_t$ that can be achieved under global unitary transformation in $\mathcal{S}_t$, that is
\begin{equation}\label{connection}
1-\mathcal{A}_{t}^d(\spst) \propto \max_{U} \mathcal{C}_d(U\rho_tU^\dagger),
\end{equation}
where the maximum is taken over all unitary matrices $U$ of dimension $t+1$, and $\mathcal{C}_d$ is the distance-based coherence monotone quantifying the coherence of $\rho_t$ in the Dicke basis~\cite{Ple14}. Equation~(\ref{connection}) provides a quantitative relation between measures of anticoherence and measure of quantum coherence. In particular, we can conclude that as the $t$-anticoherence of a spin state is greater, the less coherence at the level of its $t$-qubit reductions can be achieved in the symmetric Dicke basis.

\section{Applications}\label{sec:applications}

In this section, we use our formalism to compute various measures of anticoherence for specific spin states and to find, by numerical optimisation, anticoherent states for spin quantum numbers up to $j=10$. 

\subsection{Anticoherence measures: examples}

\subsubsection{Spin-$1$ states}

\begin{figure}[!h]
\begin{centering}
\includegraphics[width=0.45\textwidth]{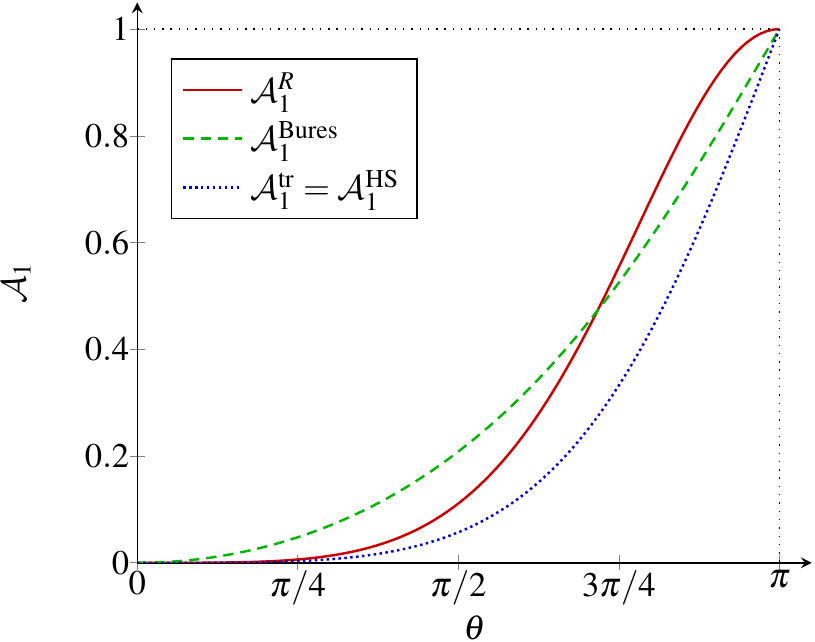}
\par\end{centering}
\caption{Measures of $1$-anticoherence (\ref{measures1ac}) for the spin-$1$ state (\ref{anyspin1}) as a function of $\theta$. \label{fig:SP1}}
\end{figure}

In the Majorana representation, any spin-$1$ state is specified by two points on the Bloch sphere that can be brought by rigid rotation in the $x-z$ plane and symmetrically opposite with respect to the $y-z$ plane. The arrangement is parametrized by the angle $\theta\in [0,\pi]$ between the lines connecting the center of the Bloch sphere to the Majorana points. The state is given in the standard basis by
\begin{equation}\label{anyspin1}
\ket{\psi_1(\theta)} = \frac{-\cot ^2\left(\frac{\theta }{4}\right)\ket{1,-1}+\ket{1,1}}{\sqrt{\cot ^4\left(\frac{\theta }{4}\right)+1}}.
\end{equation}
For $\theta=0$, the state (\ref{anyspin1}) is coherent, while for $\theta=\pi$, it is $1$-anticoherent. A direct calculation yields 
\begin{equation}\label{measures1ac}
\begin{aligned}
&\mathcal{A}_1^R = \frac{4 \sin ^4\left(\frac{\theta }{2}\right)}{(\cos \theta +3)^2},\quad\mathcal{A}_1^\mathrm{HS} =\mathcal{A}_1^\mathrm{tr} = \frac{2}{1+\cot^4\left(\frac{\theta }{4}\right)}, \\
&\mathcal{A}_1^\mathrm{Bures} = 1-\sqrt{\sqrt{2}+2-\frac{2 \sqrt{2}+2}{\sqrt{\cos \theta +3}}}.
\end{aligned}
\end{equation}
Figure~\ref{fig:SP1} shows these different measures of $1$-anticoherence as a function of $\theta$.

\subsubsection{All anticoherent states for $j=3/2$ and $j=2$}

For $j=3/2$, the only $1$-anticoherent spin state up to rotation is the Schrödinger cat state (\ref{Scatstates}). It is not $2$-anticoherent as we have $\mathcal{A}_2^R=3/4$, $\mathcal{A}_2^\mathrm{HS}=\mathcal{A}_2^\mathrm{tr}=1/2$, and $\mathcal{A}_2^\mathrm{Bures} =(1+\sqrt{2}-\sqrt{3})/2$, with all these measures being smaller than $1$.

\begin{figure}[!h]
\begin{centering}
\includegraphics[width=0.45\textwidth]{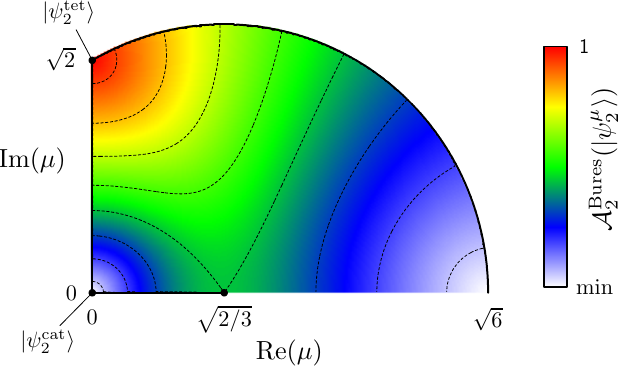}
\par\end{centering}
\caption{Density plot of the Bures measure of $2$-anticoherence [Eq.~(\ref{actBures})] of the $1$-anticoherent spin-$2$ states (\ref{eq:psimu}) as a function of the real and imaginary parts of $\mu \in \mathit{D}$ [see Eq.~(\ref{Ddef})], the only region where distinct $\mu$ define states of the form (\ref{eq:psimu}) that are not connected by a rotation. This region includes the thick black borders. The measure $\mathcal{A}^{\mathrm{Bures}}_2$ is constant along the dashed curves. The minimum value, reached for $\mu=0$, is equal to $(1+\sqrt{2}-\sqrt{3})/2$. Particular values of $\mu$ are highlighted~: $|\psi_2^{\mathrm{cat}}\rangle$ for $\mu=0$ [Eq.~(\ref{Scatstates})], $|\psi_2^{\mathrm{tet}}\rangle$ for $\mu = i\sqrt{2}$ [Eq.~(\ref{tetrahedron})], and a state which is connected by rotation to the Dicke state $ |2,0\rangle$ for $\mu =\sqrt{2/3}$.\label{fig:psimu}}
\end{figure}

For $j=2$, every $1$-anticoherent spin state can be brought by rotation to the form (\ref{eq:psimu}) with $\mu$ a $c$-number in the bounded domain~\cite{Bag14}
\begin{align}
\label{Ddef}
    \mathit{D} = \{ & \mu \in \mathbb{C} : \mathrm{Re}(\mu) \geqslant 0,\mathrm{Im}(\mu) \geqslant 0, \nonumber \\
    & |\mu - \sqrt{2/3}| \leqslant 2 \sqrt{2/3}, \mu \leqslant \sqrt{2/3} \textrm{ if } \mathrm{Im}(\mu) = 0\}
\end{align}
depicted in Fig.~\ref{fig:psimu}. The state (\ref{eq:psimu}), seen as a symmetric $4$ spin-$1/2$ state, has maximally mixed $1$ spin-$1/2$ reduced density matrices. Its $2$ spin-$1/2$ reduced density matrices in the Dicke basis all read~\cite{Bag14}
\begin{equation}\label{rho2mu}
\rho_2=\frac{1}{2+|\mu|^{2}}\left(\begin{array}{ccc}
\displaystyle
1+\frac{|\mu|^{2}}{6} & 0 & \displaystyle\sqrt{\frac{2}{3}}\,\mathrm{Re}(\mu)\\
0 & \displaystyle\frac{2}{3}|\mu|^{2} & 0\\
\displaystyle\sqrt{\frac{2}{3}}\,\mathrm{Re}(\mu)& 0 & \displaystyle 1+\frac{|\mu|^{2}}{6}
\end{array}\right),
\end{equation}
whose eigenvalues are given by
\begin{equation}\label{rho2muev}
\begin{aligned}
\lambda_1 &{}= \frac{2|\mu|^2}{3(2+|\mu|^2)}, \\
\lambda_2 &{}=\frac{6+|\mu|^2-2\sqrt{6}\,|\mathrm{Re}(\mu)|}{6(2+|\mu|^2)}, \\
\lambda_3 &{}= \frac{6+|\mu|^2+2\sqrt{6}\,|\mathrm{Re}(\mu)|}{6(2+|\mu|^2)}.
\end{aligned}
\end{equation}
From Eq.~(\ref{rho2muev}), the measures of $2$-anticoherence (\ref{actHS}), (\ref{acttr}) and (\ref{actBures}) can be easily computed for any $\mu$. Figure~\ref{fig:psimu} shows a density plot of the Bures measure of $2$-anticoherence of the state (\ref{eq:psimu}) for all $\mu\in D$~\cite{footnoteGE}. As concern the measures of $3$-anticoherence (\ref{actHS}), (\ref{acttr}) and (\ref{actBures}) of the state (\ref{eq:psimu}), they are given by $\mathcal{A}_3^\mathrm{HS}(\ket{\psi_2^\mu})=1-1/\sqrt{3}$, $\mathcal{A}_3^\mathrm{tr}(\ket{\psi_2^\mu})=1/3$, and $\mathcal{A}_3^\mathrm{Bures}(\ket{\psi_2^\mu})=1-\sqrt{2-\sqrt{2}}$ and do not depend on $\mu$. This follows from the facts that $\ket{\psi_2^\mu}$ has single spin-$1/2$ reduced density matrices $\rho_1$ with degenerated eigenvalue $1/2$ (independent of $\mu$) and that $\rho_1$ and $\rho_3$ have the same eigenvalues aside from zeros as a consequence of Schmidt decomposition.

\subsubsection{The most $2$-anticoherent state for $j=5/2$}

Because for $j=5/2$ there are no anticoherent states of order $2$~\cite{Kol08}, it is interesting to find the state that comes closest to a $2$-anticoherent state, i.e.\ to find the state with the highest measure of $2$-anticoherence. The purity, Hilbert-Schmidt and Bures measures of $2$-anticoherence were all found to be maximized by the same state,
\begin{equation}\label{s52}
\ket{\psi_{5/2}^{\rm QQ}}=\frac{1}{4}\left(-\sqrt{5}\,\ket{\tfrac{5}{2},-\tfrac{5}{2}}+\sqrt{2}\,\ket{\tfrac{5}{2},-\tfrac{1}{2}}+3\,\ket{\tfrac{5}{2},\tfrac{3}{2}}\right),
\end{equation}
for which
\begin{equation}
\begin{aligned}
& \mathcal{A}^{R}_2(\ket{\psi_{5/2}^{\rm QQ}})=\frac{99}{100},\quad \mathcal{A}^{\mathrm{HS}}_2(\ket{\psi_{5/2}^{\rm QQ}})=\frac{9}{100},\\
& \mathcal{A}^{\mathrm{Bures}}_2(\ket{\psi_{5/2}^{\rm QQ}})=1-\sqrt{\frac{-3 \sqrt{10}-\sqrt{30}+15}{15-5 \sqrt{3}}}\approx 0.9247.
\end{aligned}
\end{equation}
Interestingly, the state (\ref{s52}) coincides with the most non-classical spin state for $j=5/2$~\cite{Gir10}. However, it does not maximize all measures of $2$-anticoherence, in particular the trace measure $\mathcal{A}^\mathrm{tr}_2$. Similarly, we found that the most non-classical spin state for $j=7/2$~\cite{Gir10} is the state with the highest purity-based measure of $3$-anticoherence. However, we observed that the most non-classical spin states do not always maximize measures of anticoherence for $t=\lfloor j \rfloor$.

\subsubsection{Generalized GHZ states}

We consider the generalized Greenberger-Horne-Zeilinger (GHZ) state of $2j$ spin-$1/2$ introduced in~\cite{Dur02}, and given in the computational basis by
\begin{equation}\label{GHZgen}
\ket{\phi_S(\epsilon)} = \mathcal{N} \left( \down^{\otimes 2j} + \ket{\epsilon}^{\otimes 2j} \right)
\end{equation}
with
\begin{equation}
\ket{\epsilon} = \cos \epsilon\, \down + \sin \epsilon\, \up,
\end{equation}
where $\mathcal{N}= 1/\sqrt{2 (1+\cos^{2j} \epsilon)}$ is a normalization constant  and $\epsilon \in [0,\pi/2]$. This state allows for a continuous transition from the separable state $\down^{\otimes 2j}$ when $\epsilon =0$ to the GHZ state $(\down^{\otimes 2j} + \up^{\otimes 2j})/\sqrt{2}$ when $\epsilon =\pi/2$. As it is symmetric, it is in one-to-one correspondence with the spin-$j$ state~(\ref{eq:jm_decomp}) with
\begin{equation}\label{eq:GHZgeneralized}
\begin{aligned}
c_m = \mathcal{N} \Big [ \delta_{m,-j} + \sqrt{C_{2j}^{j+m}}  \cos^{j-m} (\epsilon)\sin^{j+m}(\epsilon) \Big ],
\end{aligned}
\end{equation}
where $\delta_{m,-j}$ is the Kronecker $\delta$.

\begin{figure}[!h]
\begin{centering}
\includegraphics[width=0.45\textwidth]{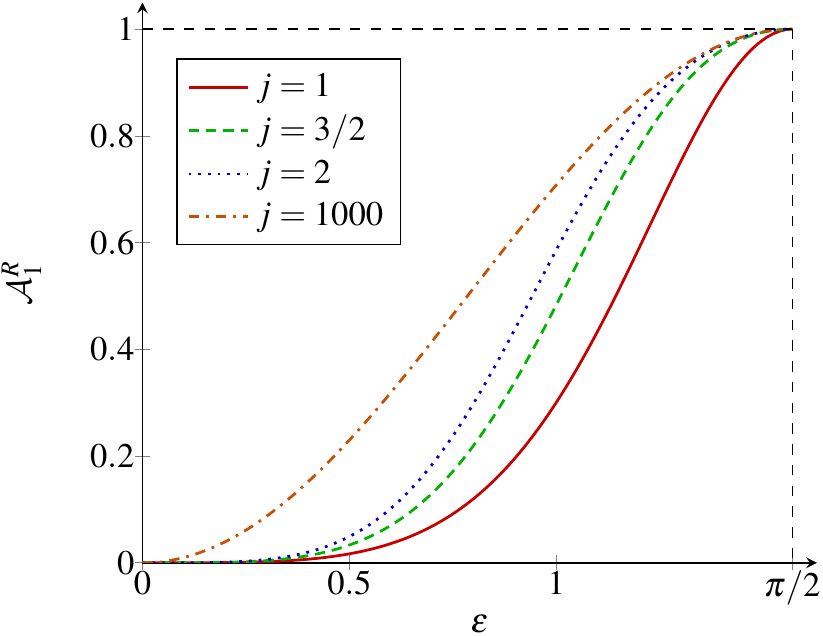}
\par\end{centering}
\caption{Purity-based measure of $1$-anticoherence (\ref{eq_acm_pure}) of the state (\ref{eq:GHZgeneralized}) as a function of $\epsilon$ for different spin quantum numbers.\label{fig:ghzgen}} 
\end{figure}

Figure~\ref{fig:ghzgen} shows the purity-based measure of $1$-anticoherence $\mathcal{A}^R_1$ as a function of $\epsilon$ for different spin quantum numbers. The fact that $\mathcal{A}^R_1$ is directly computable from the coefficients (\ref{eq:GHZgeneralized}) through Eq.~(\ref{Rtcm}) allows us to evaluate it for spin quantum numbers as large as $j=1000$. It varies continuously from $0$ for $\epsilon=0$ to $1$ for $\epsilon =\pi/2$. As the spin quantum number is larger, the transition is smoother. For $\epsilon$ close to $0$, $\mathcal{A}^R_t\approx (2j-1)\epsilon^4/4$, while for $\epsilon$ close to $\pi/2$, $\mathcal{A}^R_t\approx 1-(\epsilon-\pi/2)^2$. The generalized GHZ state (\ref{GHZgen}) serves as benchmark in the study of measures of quantum macroscopicity~\cite{Dur02}. Its macroscopicity is quantified by an effective size $N_{\mathrm{eff}}$ which scales as $N_{\mathrm{eff}}/N\approx \epsilon^2$ for $\epsilon$ close to $0$ and as $N_{\mathrm{eff}}/N\approx 1-(1+1/N)(\epsilon-\pi/2)^2$ for $\epsilon$ close to $\pi/2$, where $N=2j$ is the number of spin-$1/2$. We thus see that $\mathcal{A}^R_t$ has the same scaling with $N$ and $\epsilon$ as $N_{\mathrm{eff}}/N$ when $\ket{\phi_S(\epsilon)}$ is close to the GHZ state $\ket{\phi_S(\pi/2)}$.

\begin{figure}[!h]
\begin{centering}
\includegraphics[width=0.45\textwidth]{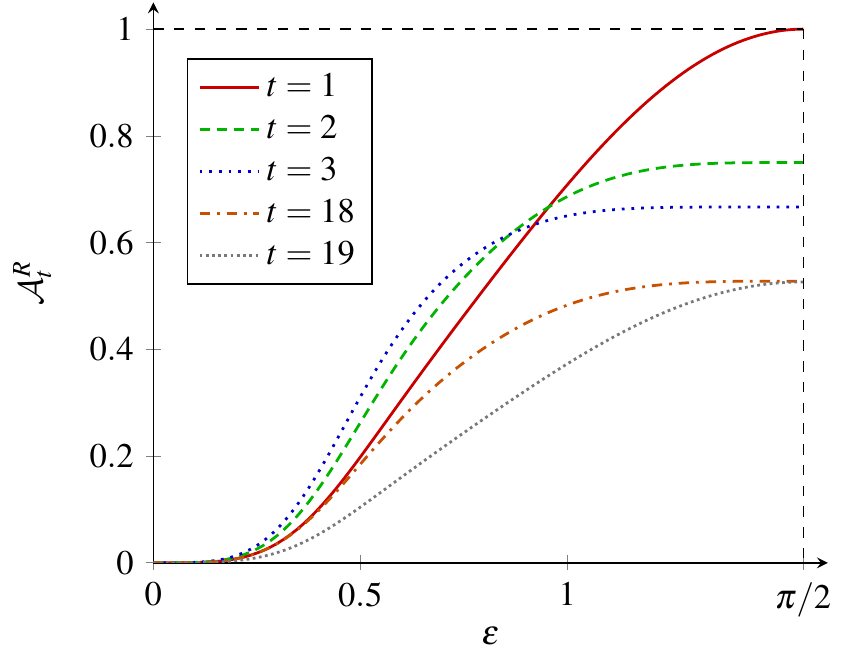}\\[6pt]
\includegraphics[width=0.45\textwidth]{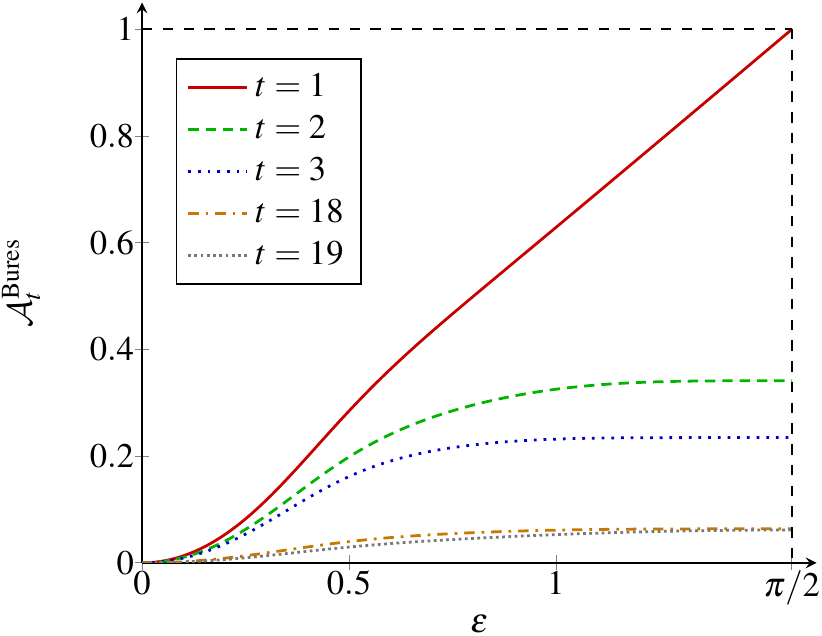}
\end{centering}
\caption{Purity-based (top) and Bures (bottom) measure of $t$-anticoherence of the state (\ref{eq:GHZgeneralized}) for $j=10$ and $t=1,2,3,18,19$ as a function of $\epsilon$.\label{fig:ghzgent}} 
\end{figure}

Figure~\ref{fig:ghzgent} shows $\mathcal{A}^R_t$ (top) and the Bures measure of anticoherence $\mathcal{A}^{\mathrm{Bures}}_t$ (bottom) as a function of $\epsilon$ for anticoherence orders $t=1,2,3,18,19$ and fixed spin quantum number $j=10$. Both figures show that the state (\ref{eq:GHZgeneralized}) is at most $1$-anticoherent because all measures with $t>1$ are strictly smaller than $1$ for any $\epsilon$. It is interesting to note that for $\mathcal{A}^{\mathrm{Bures}}_t$, the inequality~(\ref{eq:decreasing}) stating that the measure can only decrease with the order of anticoherence is verified, while for $\mathcal{A}^R_t$ an increase of the measures with $t$ is observed for some values of $\epsilon$. 

\subsubsection{Highly entangled symmetric states}

\begin{figure}[!h]
\begin{centering}
\includegraphics[width=0.45\textwidth]{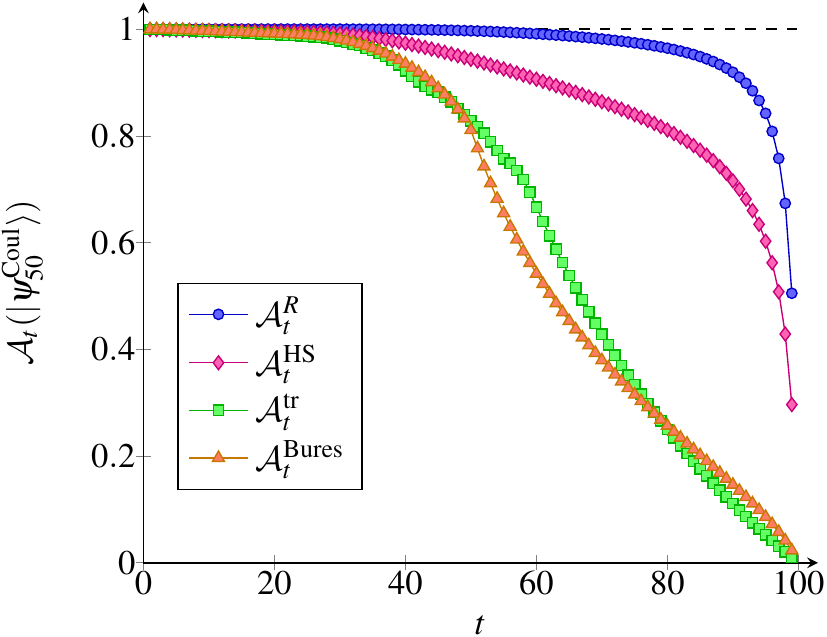}
\par\end{centering}
\caption{Measures of anticoherence as a function of the order $t$ for the spin-$50$ state $|\psi_{50}^{\mathrm{Coul}}\rangle$ whose Majorana points form an arrangement on the Bloch sphere identical to the one of point charges minimizing the Coulomb potential energy. Blue dots: $\mathcal{A}_t^R$, purple rhombus: $\mathcal{A}_t^{\mathrm{HS}}$, green squares: $\mathcal{A}_t^{\mathrm{tr}}$, and orange triangles: $\mathcal{A}_t^{\mathrm{Bures}}$. \label{psicoul}} 
\end{figure}

Highly entangled symmetric states of $2j$ spin-$1/2$ with respect to the geometric measure of entanglement can be turned into highly non-classical spin-$j$ states using the Majorana representation (see Sec.~\ref{subsec:Majo}). For large number of spins, high geometric entanglement is observed when Majorana points are spread out all over the Bloch sphere~\cite{Mar10}. One way to produce such arrangements is to consider configurations of point charges on the surface of a sphere minimizing the Coulomb potential energy, leading to symmetric states that we denote by $| \psi_j^{\mathrm{Coul}}\rangle$. Figure~\ref{psicoul} shows different measures of anticoherence as a function of the order of anticoherence computed for the state $| \psi_{50}^{\mathrm{Coul}}\rangle$. All measures are very close to $1$ for orders of anticoherence $t \lesssim 25$, meaning that $| \psi_{50}^{\mathrm{Coul}}\rangle$ is very close to a $25$-anticoherent state. Also, we observe that all measures decrease monotonously as $t$ increases. Similar results were obtained for states $| \psi_j^{\mathrm{Coul}}\rangle$ with $j<50$ and show that these states are approximately anticoherent to order $j/2$ for all $j$ considered.

\subsection{Existence of anticoherent states with degenerated Majorana points}

The existence of $t$-anticoherent spin-$j$ states has been studied in~\cite{Zimba_EJTP_3,Cra10,Bag14,Bag15,Bjo15,Per17}. It was observed that, in the Majorana representation, $t$-anticoherent states with the smallest spin quantum number correspond to arrangements of points spread out on the Bloch sphere. It was also found that $t$-anticoherent states with degenerated Majorana points exist provided $j$ is large enough~\cite{Bag15}. It should be noted that spin states for which one or several Majorana points are degenerated, such as Dicke states, play an important role in the entanglement classification under SLOCC of multiqubit symmetric states~\cite{Bas09,Aul12,Bag14}. In addition, they were shown to be useful in the design of Hardy inequalities demonstrating the persistence of non-local correlations~\cite{Mar12}. The general question of the existence of $t$-anticoherent spin-$j$ states with degenerated Majorana points can be addressed on the basis of our measures of anticoherence. Without loss of generality, we choose the most degenerated Majorana point to be at the south pole of the Bloch sphere as it can always be brought there by rotation. More specifically, the degeneracy degree $\dg$ of this point can be imposed by setting $c_m=0$ for $-j\leqslant m\leqslant -j+\dg-1$ in Eq.~(\ref{eq:jm_decomp}). Once these coefficients are set, a numerical optimization of a measure $\mathcal{A}_t$ of $t$-anticoherence can be performed on the remaining coefficients. In this work, we performed numerical optimization of the purity-based measure of anticoherence $\mathcal{A}_t^R$ as it can be computed very efficiently. We obtained results for order of anticoherence up to $t=5$ and spin quantum numbers up to $j=10$. 

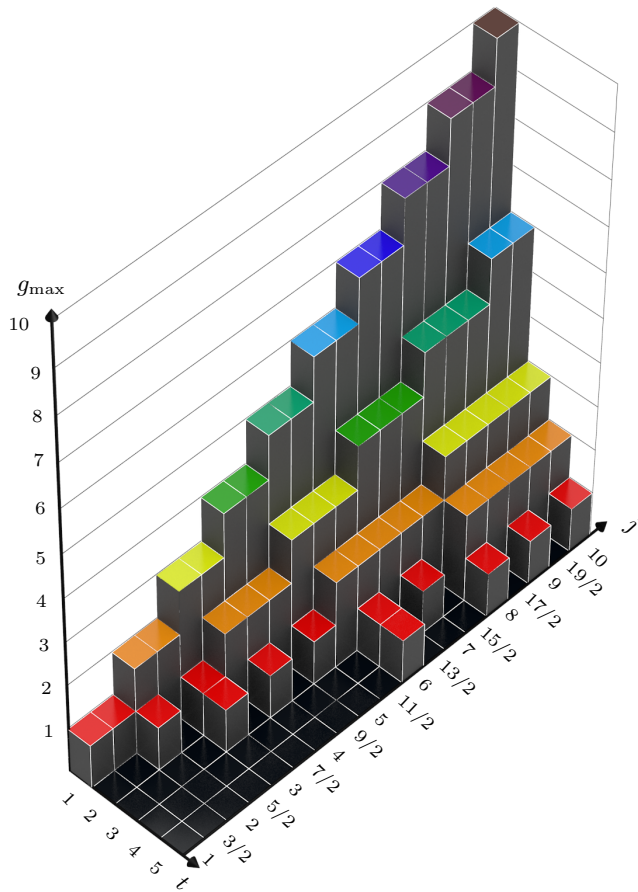
\begin{figure}[!h]
\setlength{\unitlength}{1.1mm}
\begin{overpic}[scale=0.22]{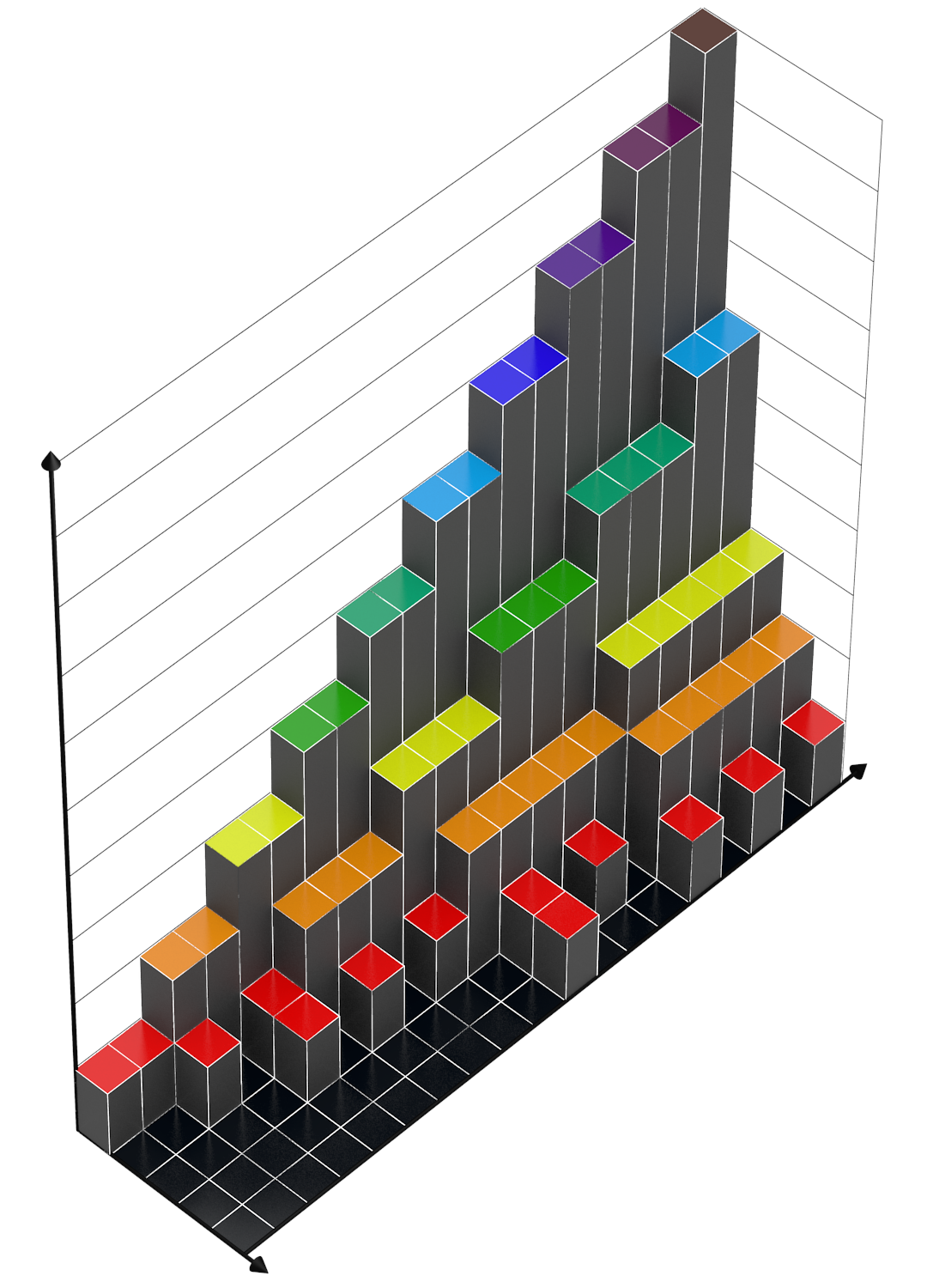}
\put(18,-1){\rotatebox{-39}{$t$}}
\put(5,9){\rotatebox{-46}{\scriptsize $1$}}
\put(7.5,7){\rotatebox{-45}{\scriptsize $2$}}
\put(10,5){\rotatebox{-44}{\scriptsize $3$}}
\put(12.5,3){\rotatebox{-43}{\scriptsize $4$}}
\put(15,1){\rotatebox{-42}{\scriptsize $5$}}

\put(0,67){$\dg_{\mathrm{max}}$}
\put(3,16){\scriptsize $1$}
\put(2.725,20.95){\scriptsize $2$}
\put(2.45,25.9){\scriptsize $3$}
\put(2.175,30.85){\scriptsize $4$}
\put(2.1,35.8){\scriptsize $5$}
\put(2.06,41.5){\scriptsize $6$}
\put(1.86,47){\scriptsize $7$}
\put(1.69,52){\scriptsize $8$}
\put(1.5,57){\scriptsize $9$}
\put(-1,62.5){\scriptsize$10$}

\put(69,40){\rotatebox{-34}{$j$}}

\put(21,2){\rotatebox{-45}{\scriptsize $1$}}
\put(23,4){\rotatebox{-43}{\scriptsize $3/2$}}
\put(26,6){\rotatebox{-45}{\scriptsize$2$}}
\put(28,8){\rotatebox{-43}{\scriptsize$5/2$}}
\put(30.9,10){\rotatebox{-47}{\scriptsize$3$}}
\put(33.1,12){\rotatebox{-43}{\scriptsize$7/2$}}
\put(35.8,14){\rotatebox{-47}{\scriptsize$4$}}
\put(38,16){\rotatebox{-43}{\scriptsize$9/2$}}
\put(40.7,18){\rotatebox{-47}{\scriptsize$5$}}
\put(43,20){\rotatebox{-43}{\scriptsize$11/2$}}
\put(45.7,22){\rotatebox{-47}{\scriptsize$6$}}
\put(47.7,24){\rotatebox{-43}{\scriptsize$13/2$}}
\put(50.6,26){\rotatebox{-47}{\scriptsize$7$}}
\put(52.4,27.8){\rotatebox{-43}{\scriptsize$15/2$}}
\put(55.6,30){\rotatebox{-47}{\scriptsize$8$}}
\put(57.5,31.7){\rotatebox{-43}{\scriptsize$17/2$}}
\put(60.5,33.5){\rotatebox{-47}{\scriptsize$9$}}
\put(62.2,35.3){\rotatebox{-43}{\scriptsize$19/2$}}
\put(65,37){\rotatebox{-47}{\scriptsize$10$}}
\end{overpic}

\protect\caption{Largest degeneracy degree $\dg_{\mathrm{max}}$ of at least one of the Majorana points for which $t$-anticoherent states with a spin quantum number $j$ are found numerically by optimization of the purity-based measure of anticoherence $\mathcal{A}_t^{R}$. When no states are found, we set $\dg_{\mathrm{max}}=0$. In particular, the minimal spin quantum number $j_{\mathrm{min}}$ for which $t$-anticoherent states are found can be read from this figure: $j_{\mathrm{min}}=1$ for $t=1$, $j_{\mathrm{min}}=2$ for $t=2$, $j_{\mathrm{min}}=3$ for $t=3$, $j_{\mathrm{min}}=6$ for $t=4$ and $j_{\mathrm{min}}=6$ for $t=5$. 
\label{maxdegeneracy}}
\end{figure}

Figure~\ref{maxdegeneracy} shows the largest degeneracy degree $\dg_{\mathrm{max}}$ among the Majorana points allowing for the existence of $t$-anticoherent spin-$j$ states. Each block corresponds to a state for which optimization of $\mathcal{A}_t^R$ yields $|\mathcal{A}_t^R-1|<10^{-10}$. The block height gives the degeneracy degree $\dg_{\mathrm{max}}$ of the most degenerated Majorana point of that state. The absence of block indicates the lack of convergence towards a $t$-anticoherent state. From this figure, the existence of $t$-anticoherent spin-$j$ states for a given couple $(t,j)$ can be read off.  It is interesting to note that for the range of values of $t$ considered, the state with the smallest $j$ found has always non-degenerated Majorana points. We also observe that the minimal $j$ for the existence of a $t$-anticoherent state with a maximal degeneracy degree $\dg_{\mathrm{max}}$ is always smaller than the minimal $j$ for the existence of a $t$-anticoherent state with a larger maximal degeneracy degree $\dg_{\mathrm{max}}+1$. Last, we see that the existence of a $t$-anticoherent state of spin quantum number $j$ does not imply the existence of $t$-anticoherent states for all spin quantum number larger than $j$. In the following, we list some of the states that we found or deduced from the results of our numerical optimization.

\begin{figure*}
\begin{overpic}[width=0.95\textwidth,clip=true]{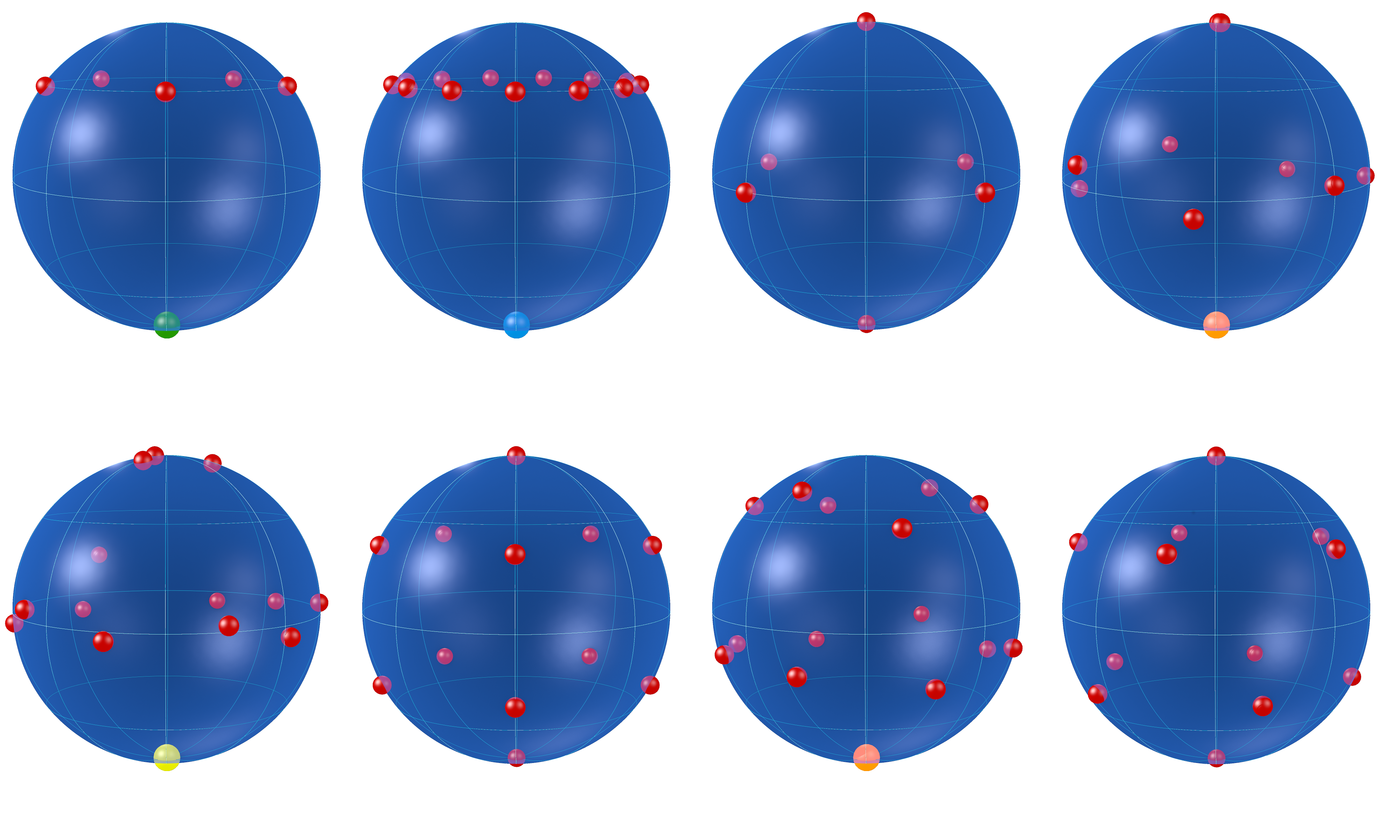}
\setlength{\unitlength}{1mm}
\put(5,56){\small $\begin{array}{c}
t=1\\
\dg=4\\
\end{array}$, $j_{\mathrm{min}}=9/2$}
\put(47,56){\small $\begin{array}{c}
t=2\\
\dg=6\\
\end{array}$, $j_{\mathrm{min}}=19/2$}
\put(93,56){\small $\begin{array}{c}
t=3\\
\dg=1\\
\end{array}$, $j_{\mathrm{min}}=3$}
\put(133,56){\small $\begin{array}{c}
t=3\\
\dg=2\\
\end{array}$, $j_{\mathrm{min}}=11/2$}

\put(7,3){\small $\begin{array}{c}
t=3\\
\dg=3\\
\end{array}$, $j_{\mathrm{min}}=8$}
\put(50,3){\small $\begin{array}{c}
t=4\\
\dg=1\\
\end{array}$, $j_{\mathrm{min}}=6$}
\put(93,3){\small $\begin{array}{c}
t=4\\
\dg=2\\
\end{array}$, $j_{\mathrm{min}}=8$}
\put(136,3){\small $\begin{array}{c}
t=5\\
\dg=1\\
\end{array}$, $j_{\mathrm{min}}=6$}
\end{overpic}

\caption{Majorana representation of states with minimal spin quantum number $j_{\mathrm{min}}$ for a given order $t$ of anticoherence and a given degeneracy degree $\dg$ of one of the Majorana points (the point located at the South pole). From top left to bottom right: (\ref{eq:t1dmaxnoninteger}) with $j=9/2$, (\ref{eq:maxdegt2}) with $j=19/2$, (\ref{octstate}), (\ref{eq:n11t3d2}), (\ref{eq:n16t3d3}), (\ref{eq:icorot}) with $\theta=0$, and (\ref{eq:n16t4d2}), (\ref{eq:icorot}) with $\theta=\pi/2$. The small dots represent non-degenerated Majorana points, while the large dots represent Majorana points with $\dg$-fold degeneracy as specified in the labels (we use the same color code as in Fig.~\ref{maxdegeneracy}). \label{allspheres}}
\end{figure*}
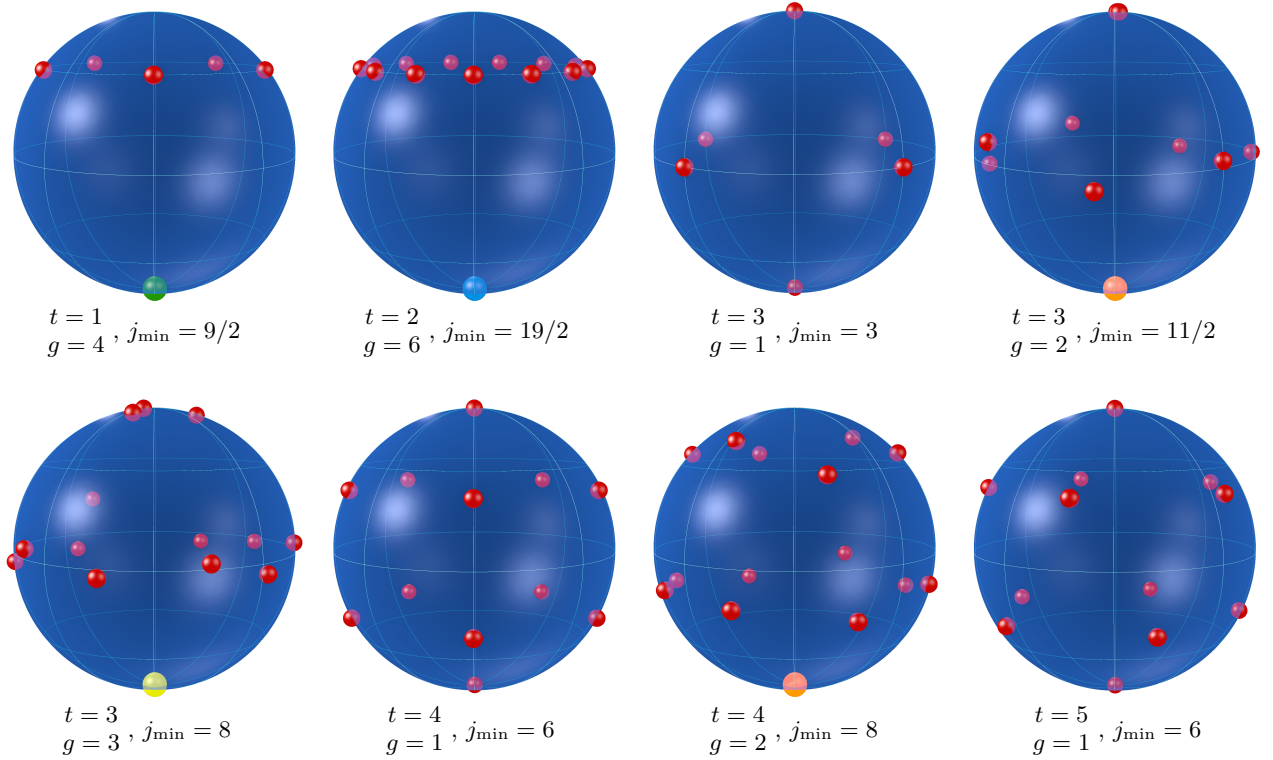

\subsubsection{$1$-anticoherent states}

There are $1$-anticoherent states for any $j>1/2$. The maximal degeneracy degree of their Majorana points was found to depend on the parity of $2j$. For integer $j$, $\dg_{\mathrm{max}} = j$ and the corresponding states are (up to rotation)
\begin{equation}\label{eq:t1dmaxinteger}
\spst=\ket{j,0}.
\end{equation}
For half-integer $j$, $\dg_{\mathrm{max}}  = j-1/2$ and the corresponding states are (up to rotation)
\begin{equation}\label{eq:t1dmaxnoninteger}
\spst = \frac{1}{\sqrt{2j+1}}(\sqrt{2j}\,\ket{j,-\tfrac{1}{2}}+\ket{j,j}).
\end{equation}
The Majorana representation of the states (\ref{eq:t1dmaxinteger}) corresponds to $j$ points at the south pole and $j$ points at the north pole of the Bloch sphere, while for the states (\ref{eq:t1dmaxnoninteger}) it corresponds to one $(j-1/2)$-fold degenerated point at the south pole of the Bloch sphere and $(j+1/2)$ non-degenerated points lying at the apex of a regular polygon parallel to the equator in the northern hemisphere. This latter arrangement is illustrated for $j=9/2$ in Fig.~\ref{allspheres} (top left). 

\subsubsection{$2$-anticoherent states}
The smallest spin quantum number for which 2-anticoherent spin states with $\dg$-fold degenerated Majorana points were found numerically is $j_{\dg} = (1+3\,\dg)/2=2,7/2,5,\ldots,19/2$. An example of such state for all $j_{\dg}$ of the form $(1+3\,\dg)/2$ with integer $\dg$ is given by
\begin{equation}\label{eq:maxdegt2}
\ket{\psi_{j_{\dg}}}=\sqrt{\frac{3\,j_{\dg}}{4j_{\dg}+1}}\,\left|j_{\dg},-\frac{j_{\dg}+1}{3}\right\rangle+\sqrt{\frac{j_{\dg}+1}{4j_{\dg}+1}}\ket{j_{\dg},j_{\dg}}.
\end{equation}
Its Majorana representation is made of a $\dg$-fold degenerated point at the South pole and non-degenerated points lying at the apex of a regular $(N-\dg)$-gone parallel to the equator in the Northern hemisphere. This arrangement of points for $\dg = 6$ ($j_\textrm{min} = 19/2$) is shown in Fig.~\ref{allspheres}. In particular, for $\dg = 1$, the state (\ref{eq:maxdegt2}) is connected by a rotation to the spin-$2$ tetrahedron state (\ref{tetrahedron}). Note that it was shown in~\cite{Cra10} that any state with a tetrahedral symmetry of its Majorana points is 2-anticoherent. While there exists a $2$-anticoherent state for $j=2,3$, Fig.~\ref{maxdegeneracy} shows that there is no $2$-anticoherent state for $j=5/2$ (hole in the $t=2$ line).

\subsubsection{$3$-anticoherent states}

The smallest spin quantum number for which 3-anticoherent spin states without degenerated Majorana points were found numerically is $j=3$. Their Majorana representation corresponds to points at the apex of an octahedron (octahedral symmetry is known to imply anticoherence to order 3~\cite{Cra10}), see Fig.~\ref{allspheres}. These states can be brought by rotation to the form
\begin{equation}\label{octstate}
\ket{\psi_3^{\textrm{oct}}} = \frac{1}{\sqrt{2}}\left(\ket{3,-2}+\ket{3,2}\right).
\end{equation}

The smallest spin quantum number for which 3-anticoherent spin states with $2$-fold degenerated Majorana points were found numerically is $j=11/2$. An example of such state is given in the appendix by Eq.~(\ref{eq:n11t3d2}). Its Majorana representation is shown in Fig.~\ref{allspheres} (top right).

The smallest spin quantum number for which 3-anticoherent spin states with $3$-fold degenerated Majorana points were found numerically is $j=8$. An example of such state is given in the appendix by Eq.~(\ref{eq:n16t3d3}). Its Majorana representation is shown in Fig.~\ref{allspheres} (bottom left).

\subsubsection{$4$-anticoherent states}
The smallest spin quantum number for which 4-anticoherent spin states without degenerated Majorana points are found numerically is $j=6$. A family of such states not connected by rotations is given by
\begin{equation}\label{eq:icorot}
\ket{\psi_6(\theta)}= \frac{1}{5}(
\sqrt{7}\,\ket{6,-5}+
\sqrt{11}\,e^{i\theta}\, \ket{6,0}+
\sqrt{7}\,\ket{6,5})
\end{equation}
with $0\leqslant \theta < \pi/2$. The Majorana representation of (\ref{eq:icorot}) with $\theta=0$ is shown in Fig.~\ref{allspheres}.

The smallest spin quantum number for which 4-anticoherent spin states with twofold-degenerated Majorana points are found numerically is $j=8$. An example of such state is given in the appendix by Eq.~(\ref{eq:n16t4d2}). Its Majorana representation is shown in Fig.~\ref{allspheres}.

\subsubsection{$5$-anticoherent states}
The smallest spin quantum number for which 5-anticoherent spin states without degenerated Majorana points are found numerically is $j=6$. An example of such state is given by (\ref{eq:icorot}) with $\theta=\pi/2$. Its arrangement of Majorana points displays icosahedral symmetry (see Fig.~\ref{allspheres}, bottom right), which implies anticoherence to order 5~\cite{Cra10}.

\section{Conclusion and perspectives}

In this paper, we have introduced the notion of measure of anticoherence to order $t$ for pure spin states with arbitrary spin quantum number $j$. By definition, these  measures allow us to position any pure spin state on a scale ranging from 0 --only for coherent states-- to 1 --only for $t$-anticoherent states. By exploiting the one-to-one correspondence between spin-$j$ states and symmetric states of $2j$ spin-$1/2$ (Majorana representation), we have devised a measure of $t$-anticoherence for spin-$j$ states based on the purity of its reduced density matrices $\rho_t$ of $t$ spin-$1/2$. In particular, our purity-based measure of anticoherence reduces to a linear function of the total variance in the case $t=1$. We then have presented a general method to construct measures of anticoherence based on operator distances. All these measures can be directly extended to mixed states as long as $t<2j$. While for mixed states the order $t=2j$ is relevant (the maximally mixed state of spin $j$ is $2j$-anticoherent), our constructions based on reduced density matrices no longer apply. We have exemplified this method with the Hilbert-Schmidt, the trace and the Bures distances, and have discussed the relation of these distance-based measure of anticoherence with measures of quantum coherence. All our measures have the practical advantage of being easily computable because they do not require any optimization over a set of states. In particular, a closed form expression for the purity-based measure of $t$-anticoherence has been obtained. As for the distance-based measures of $t$-anticoherence, they have been expressed as simple functions of the eigenvalues of $\rho_t$. As an illustration of our measures, we have calculated their value for arbitrary spin-$1$ states, all $1$-anticoherent states of spin $j=3/2$ and $j=2$, and states with higher spin quantum numbers, such as generalized GHZ states or states with high geometric entanglement. We also have used our measures to study the problem of the existence of $t$-anticoherent spin-$j$ states with degenerated Majorana points for order of anticoherence up to $t=5$ and spin quantum numbers up to $j=10$. Our results reveal the intricate link between degeneracy of Majorana points and anticoherence.

A direct extension of this work concerns the design of measures of polarization for multiphoton states (or degrees of quantum polarization, see, e.g.,~\cite{Mar10,Hoz13}). By identifying the spin operators to the Stokes operators, our formalism presented here for spin states can be directly transposed to multiphoton states. In this perspective, the Bures measure of anticoherence appears particularly appropriate as it enjoys the property of monotonicity in the order $t$ of anticoherence. Another possible direction of investigation concerns quantum metrology. As anticoherent states have been shown to be optimal in detecting rotations~\cite{Chr17} and for reference frame alignment~\cite{Kol08}, it would be worth investigating the connections between our measures of anticoherence and the efficiency of a state for such tasks.

\acknowledgments
Computational resources were provided by the Consortium des Equipements de Calcul Intensif (CECI), funded by the Fonds de la Recherche Scientifique de Belgique (F.R.S.-FNRS) under Grant No. 2.5020.11.

\appendix

\section*{Appendix: Some anticoherent states with degenerated Majorana points}

In this appendix, we list some anticoherent states found numerically with the smallest spin quantum number for a given order $t$ and a given degeneracy degree $g$ of one of their Majorana points (see Sec.~\ref{sec:applications} and Fig.~\ref{allspheres}). These spin-$j$ states are given in terms of their expansion coefficients in the Dicke basis $(c_{-j}\:c_{-j+1}\:\ldots\:c_{j-1}\:c_j)^T$ [see Eq.~(\ref{eq:jm_decomp})].

\subsubsection{$t=3$, $\dg=2$, $j_{\mathrm{min}}=11/2$}
{\footnotesize
\begin{equation}
\label{eq:n11t3d2}
\left(
\begin{array}{c}
0\\
0\\
0.6189711605133+0.3210948626046\, i\\
0.0035795645781-0.005571932846\, i\\
0.0000596970141+0.0009612745249\, i\\
0.0747280614210+0.0848752159787\, i\\
-0.098250832667+0.0704276863999\, i\\
-0.004358698832-0.006121053115\, i\\
0.0169591633687+0.0449205206870\, i\\
0.6727762527486-0.173404352179\, i\\
0.0053207161522+0.0351899547234\, i\\
-0.001014420524-0.000272398051\, i
\end{array}
\right)
\end{equation}
}

\subsubsection{$t=3$, $\dg=3$, $j_{\mathrm{min}}=8$}
{\footnotesize
\begin{equation}
\label{eq:n16t3d3}
\left(
\begin{array}{c}
0\\
0\\
0\\
0.6207434617909+0.3092681061476\, i\\
-0.004351945720-0.004576402817\, i\\
0.0012063346305-0.004493986067\, i\\
-0.018457273316+0.0463722998675\, i\\
0.0655377989379+0.0201067990800\, i\\
0.0686716910441-0.011023764770\, i\\
0.0455872510982+0.1357843214759\, i\\
-0.033716686148+0.0740640065423\, i\\
-0.065020180326+0.0699845281978\, i\\
-0.142220507502+0.0527191543731\, i\\
0.6344068714556+0.1721745869811\, i\\
0.0530094546887+0.0724148358782\, i\\
0.0113869490780+0.0848466671314\, i\\
0.0127861473227+0.0031452746268\, i
\end{array}
\right)
\end{equation}
}

\subsubsection{$t=4$, $\dg=2$, $j_{\mathrm{min}}=8$}
{\footnotesize
\begin{equation} 
\label{eq:n16t4d2}
\left(
\begin{array}{c}
0\\
0\\
0.3232497765551+0.4980926832112\, i\\
-0.002755440315+0.0002941675004\, i\\
0.0096608735602-0.019233596605\, i\\
0.0353301997743+0.0318247115315\, i\\
0.0938165016555-0.001235092383\, i\\
-0.003767421017-0.082840446425\, i\\
0.0895251593971-0.005880000805\, i\\
0.0127309067916-0.038624872627\, i\\
0.2264247580540+0.6299063613884\, i\\
0.0268965215414+0.0274972703211\, i\\
0.0799844343901-0.093411408577\, i\\
0.0206511586120+0.0431241880491\, i\\
0.0456490431434-0.141531955144\, i\\
0.0053521006629-0.007262562142\, i\\
0.3557695532332+0.0599218154303\, i
\end{array}
\right)
\end{equation}
}


\begin{thebibliography}{99}
\bibitem{footnote1} This state is also known in the literature as the tetrahedron state since in the Majorana representation, its defining four Majorana points draw a regular tetrahedron.
\bibitem{footnoteGE} Note that Fig.~3 in~\cite{Bag14} shows the geometric entanglement of the states (\ref{eq:psimu}) as a function of $\mu$. The geometric entanglement is related to the Fubini-Study distance of the states (\ref{eq:psimu}) to the set of separable states, whereas the Bures measure of $2$-anticoherence is related to the Bures distance between the $2$ spin-$1/2$ reduced state and the maximally mixed state. Interestingly, there is a certain degree of similarity between this figure and Fig.~\ref{fig:psimu} of the present paper.
\end{thebibliography}
\end{document}